# The determinant factors for model resolutions obtained using CryoEM method

Yihua Wang[1], Daqi Yu[1], Qi Ouyang[1], and Haiguang Liu[2*]

[1]*Key Laboratory for Artificial Microstructure and Mesoscopic Physics, Institute of Condensed Matter Physics, School of Physics, Center for Quantitative Biology School of Physics, The Peking-Tsinghua Center for Life Sciences at School of Physics, Peking University, Beijing, China*

[2]*Complex Systems Division, Beijing Computational Science Research Centre, Beijing, China 100193*

[*] Corresponding author, Email: hgliu@csrc.ac.cn

**Abstract**

The CryoEM single particle imaging method has recently received broad attention in the field of structural biology for determining the structures of biological molecules. The structures can be resolved to near-atomic resolutions after rending a large number of CryoEM images measuring molecules in different orientations. However, the factors for model resolution need to be further explored. Here, we provide a theoretical framework in conjunction with numerical simulations to gauge the influence of several key factors that are determinant in model resolution. We found that the number of measured projection images and the quality of each measurement (quantified using average signal-noise-ratio) can be combined to a single factor, which is dominant to the constructed model resolution. Furthermore, the intrinsic thermal motion of the molecules and the defocus levels of the electron microscope both have significant effects on the model resolution. These effects can be quantitatively summarized using an analytical formula that provides a theoretical guideline on structure resolutions for given experimental measurements.

**Highlights:**

- Using numerical simulation methods to evaluate the key factors that affect the resolution of CryoEM structures within a theoretical framework;

- Incorporating the intrinsic thermal fluctuation of the molecules in the analytical formula to better describe the limiting factor of the resolution due to structural heterogeneity;

- Providing a theoretical guideline on structure resolutions for given experimental parameters.

1. Introduction

The Cryo-electron microscopy (CryoEM) single particle imaging method has become popular recently in the structural biology research community (Bai., McMullan., and Scheres. 2015). The basic idea of Cryo-EM is to measure the particles at all possible orientations and use model reconstruction algorithms to build a 3D structure that best satisfies the overall measurements. Due to irradiation damage from high-energy electrons during the measurements, each particle or molecule can only tolerate certain electron doses before the molecule is deteriorated. Experimentally, each particle/molecule is only measured once at a given orientation that is nearly fixed in vitreous ice. The CryoEM spreads the total dosage to a large ensemble of molecules, and each scatters a tolerable number of electrons to form a magnified image. Meanwhile, the cryogenic environment protects the sample molecules, maintaining molecular integrity. Nonetheless, the model resolutions from CryoEM experiments were not close to those obtained from the X-ray crystallography method until the recent breakthrough. This breakthrough includes three aspects, namely (1) the invention of a direct electron detector to allow accurate and fast measurement of electrons (Faruqi and McMullan 2011); (2) the development of data processing software, in particular the application of Bayesian algorithms in reconstructions, backed by high-performance computers (Tang et al. 2007; Scheres 2012; Grigorieff 2016); and (3) the advances in sample preparations that allow measurements at diverse orientations using thin vitreous ice layers (da Fonseca and Morris 2015; Passmore and Russo 2016; Bernecky et al. 2016). The fast readout rate of the new camera also enables measurements in movie modes that lead to the correction of molecular drift during data collection to sharpen the blurred signals (Li et al. 2013; Zheng et al. 2017). Since the resolution breakthrough reported in the structural determination of the TRPV1 molecule at 3.4 Å (Cao et al. 2013), many high-resolution structures of molecular complexes have been determined using the CryoEM single particle imaging method. The technology is enriching in the protein structure database, particularly with large molecular complexes (Bernstein et al. 1977; Berman et al. 2000; Newman et al. 2002; Newman, Tagari, and Chagoyen 2003; Rose et al. 2015; Nogales 2015; Lawson et al. 2016).

Given all the advances in CryoEM single particle imaging method, there are fundamental questions remained. One question we would like to address here is regarding the determinant factors for model resolution. In the crystallography method, the concept and measures of resolution have been well established (Morris et al. 1992; Karplus and Diederichs 2012), while they are still under investigation in CryoEM. In general, the resolutions for the structures determined using the CryoEM approach are estimated using model consistency, i.e., by calculating the Fourier shell correlation profiles and examining the point where the signal disappears (Heel and Harauz 1986; Böttcher, Wynne, and Crowther 1997; Rosenthal and Henderson 2003; Scheres and Chen 2012). Recently, several alternative methods have been developed to assess the model resolution, such as the approach that checks the local details of the structure (Kucukelbir, Sigworth, and Tagare 2013). Regardless of the different definitions in the resolution, the correct interpretation of the reconstructed structure is subject to validation using complementary approaches, such as biochemical assays or fluorescence single molecule experiments. Putting aside the arguments on CryoEM resolutions using different approaches, we would like to focus on the factors that determine the model resolution and hope to obtain a theoretical framework that guides the experiments to improve the resolution using optimized protocols for data collection.

In this work, we investigated four factors that influence the model resolution, namely, the number of projection measurements, the signal-to-noise ratio (SNR) levels of individual measurements, the electron microscope defocus, and the intrinsic flexibility of the molecules. Early studies have provided important clues about how these factors may contribute to the model resolutions. For example, Henderson studied the resolution limits that resulted from electron microscopy and provided a relation between the resolution and the number of projections (Henderson 1995). Later, Rosenthal and Henderson formulated a more detailed equation (RH model) to estimate the desired number of projections for the different structure resolutions (Rosenthal and Henderson 2003). In the RH model, the electron dose, SNR, molecular symmetry, and an effective B-factor were considered (Rosenthal and Henderson 2003; Penczek 2010; Liao and Frank 2010). The effective B-factor was found to be important, as it is used to model the combined effects of molecular drifting due to charging effects, molecular flexibility, errors in image processing, etc., into a Gaussian envelope function that describes the signal falloff (Liao and Frank 2010; Penczek 2010). Based on this pioneer research, we would like to revisit these relations and validate the formulations using numerical simulations. On the other hand, it is known that many molecules undergo dynamic motions to be functional. To solve structures at higher resolutions, the molecules can be locked in a particular conformation. For example, Subramaniam and coworkers used a cell-permeant inhibitor to stabilize β-galactosidase and obtained a CryoEM structure at 2.2 Å (Alberto et al. 2015). In other work, they obtained a structure of glutamate dehydrogenase to 1.8 Å after detailed projection classifications by sorting out the images that belong to the most populated conformation (Merk et al. 2016). We set out to investigate the effects of molecular intrinsic motion using a structure ensemble to simulate the CryoEM images, taking the heterogeneous conformation reality into consideration. Finally, electron microscopes were operated in defocus mode to obtain the contrast, and we found that the defocus levels also influenced the resulted model resolutions. Here, we proposed a framework using these factors to predict structure the resolutions. The numerical simulation results are used to estimate the free parameters. The statistics from the resolved structures are consistent with the proposed model.

## 2. Method and Theory

### 2.1 Existing theoretical framework

The model initially proposed by Rosenthal and Henderson (the RH model) and later explained by Liao and Frank connects key elements in the CryoEM method using the following equation:

$$N(k) = \frac{C\sigma^2}{1/n_k \sum_{r=1}^{n_k} |F_T^r(k)|^2} k e^{\frac{Bk^2}{2}} \qquad (1)$$

where $N$ is number of particles (projections) that are needed to reach the resolution defined by the Fourier frequency $k$; $\sigma^2$ is the variance of the noise, $F_T^r(k)$ is the Fourier intensity of the structure in the resolution shell $[k,k+\Delta k]$, $B$ is the temperature factor, and $C$ is a scaling constant.

The signal-to-noise ratio at a resolution shell $k$ is effectively represented by $\frac{1/n_k \sum_{r=1}^{n_k} |F_T^r|^2}{\sigma^2}$, where the numerator resembles the rotational average of the intensity (similar to the small/wide angle scattering intensity) and the denominator is the noise level. The second term on the right side of the equation, $k$, describes the linear dependency of $N(k)$ on the resolution shell $k$ because the number of data points in each 2D projection is at the order of $k^2$, while the desired number of data points in 3D increases with $k^3$. The last term is the Gaussian falloff to account factors, such as structural fluctuations or misalignment during data processing. We carried out numerical simulations to examine the effects of each component. Based on the results, we revise the formula to provide an improved prediction formula on model resolutions.

## 2.2 Data simulation

The structure of GroEL (PDB ID: 1XCK, (Bartolucci. et al. 2005)) was used as the model system in the numerical simulations (see Figure 1). We present two models to investigate the factors that influence the structure resolutions, the Gaussian Noise (GN) model and the Thermal Fluctuation (TF) model, as shown in Figure 1.

The Gaussian noise (GN) model was proposed to describe the relationship between the resolution and the number of projections ($N_p$), the variance of the Gaussian noise, and the defocus levels.

$$N(k) = A_{GN} \sigma^2 k e^{Bk^2} \qquad (2)$$

where $N(k)$ is the desired number of particles (projections) to collect confident signal levels at a frequency k; $\sigma^2$ is the variance of the Gaussian noise; and $B$ is a scaling parameter.

The TF model describes how thermal fluctuations of the molecules influence the structure resolution. RMSD (root-mean-square-deviation) is one quantity that measures the structural difference between the models, and here we used the mean square of the pairwise RMSD in an ensemble of structures to measure the structure fluctuations and to mimic the Debye–Waller factor. The following formula is proposed to relate the thermal fluctuation levels, number of projection images and achievable structural resolution:

$$N(k) = A_{TF} k e^{C \langle RMSD^2 \rangle k^2} \qquad (3)$$

where $\langle RMSD^2 \rangle$ is the mean square of the RMSD values obtained by pairwise comparison within the structure ensemble; and $k$ is the spatial frequency. In both equations (2) and (3), $A_{GN}$, $A_{TF}$, $B$, and $C$ are free parameters.

For the Gaussian noise model, SPIDER 22.03 (Shaikh et al. 2008) packages were used to generate the simulation data. The atomic model of GroEL was first converted to a density map (voxel size=(0.86 Å)$^3$), and then projection images were simulated at orientations generated with the successive orthogonal rotation sample approach (Yershova et al. 2010). The noises were incorporated according to the desired signal-to-noise ratio following a Gaussian distribution after the contrast transfer function (CTF) was added to the simulated projections. The programs in the SPIDER package were used through the simulation process, which is summarized in Figure 2a.

To simulate the heterogeneity in the structures, we first obtained diverse structures to form a structure ensemble based on the GroEL structure. Without the loss of generality, the structural ensemble was generated using the normal mode perturbation approach. The 'ProDy' program based on an anisotropic elastic network model was used to compute the normal modes and the eigenvalue spectrum (Bakan et al. 2014). Since the functional relevant motions are highly collective, the three models corresponding to the lowest frequencies were used to compute the perturbed structures. Specifically, the original structure (gray colored in Figure 1b) was perturbed along the three normal modes (accounting to approximately 20.8% fluctuations), with deformation amplitudes ranging from 1 to 100, and an ensemble of 1000 structures was generated around the original structure, with an RMSD of up to approximately 10 Å. The RMSD values of the generated structure relative to the original structure were used to group the structures into 10 bins using 1 Å as the bin size. The structure ensembles were then compiled by drawing structures from the 10 bins using the following approach: the first group includes structures randomly drawn from the first bin (RMSD<1 Å compared to the original structure). The second group is composed of structures randomly drawn from the first two bins, etc. This is to ensure that the structure diversities are different in the 10 groups. This procedure is summarized in Figure 2b. The average structure deviation from the original structure, as well as the mean square of the pairwise RMSD within each group, which is used to measure the structure diversity, is summarized in Table 1. This allows us to examine the dependence of the model resolution on the structure diversity (or thermal fluctuation levels).

The projection image simulation procedure is essentially the same as previously described for the case of a single 3D structure, except that for each image a 3D structure was randomly chosen from the corresponding group. As a result, each simulated dataset has the following controlled parameters: the number of projections, signal-to-noise ratio, defocus levels, and the structure heterogeneity due to the thermal fluctuations.

2.3 Resolution cutoff and curve fitting

The resolution determination was based on the Fourier shell correlation (FSC) implemented in SPIDER with the Golden Standard rule at the cutoff level of FSC = 0.143. The simulation data was split into two half subsets randomly, and each reconstruction was carried out using back projections with known orientations using the Relion 1.4 package (Scheres 2012).

The resolutions of the reconstructed models were determined for various combinations of factors that were considered in this work. The value of each factor was systematically scanned in practical ranges so that the quantitative relationships could be studied using a parameter fitting approach to the theoretical models. The free parameters used in the GN and TF models were obtained by the nonlinear Least Squares (Curve Fitting) method in MATLAB.

2.4 The survey of resolutions of experimental models

A survey was carried out on the structures determined using CryoEM deposited in the EMDB database (http://www.ebi.ac.uk/pdbe/). The retrieved information includes the molecular weight, detector type, resolution determination criteria, molecular symmetry, number of projections and the microscope operating voltages.

We focused on the models that fulfill the following requirements: (1) Models that were deposited between 2014/01/01 and 2016/04/25; (2) Micrographs were recorded using direct electron detection technology; (3) The reported resolution is determined with golden standard rule at FSC=0.143; and (4) Models without higher symmetry. As a result, 59 models were selected to for the resolution statistics (See Supplementary file for a complete list of the models).

3. Results

The RH model describes the dependency of the resolution on the number of projections and the signal-to-noise-ratio, molecular symmetry, and other factors under the umbrella of the b-factor. We focused on the study of four parameters by simplifying the formula to the GN and TF models, as described in the Methods section. By varying the parameters that quantify these determinant factors, the influence on the resolution of the reconstructed structures were systematically evaluated. The model system datasets were used to estimate the free parameters by optimizing the fit of the theoretical formula to the data points obtained from the simulations.

### 3.1. Effective number of projections

The RH model describes the dependency of structure resolution on the number of experimental images, and the logarithm trend of the reconstructed model resolution as a function of the number of measured projections is attributed to the b-factors (due to sample particle drifting, misalignment, numerical interpolation, etc.) (Rosenthal and Henderson 2003). Surprisingly, the logarithm trend was observed for the simulation data without explicitly applying the envelope function $e^{\frac{Bk^2}{2}}$ during the projection simulations. Furthermore, the utilization of the known orientation as available information eliminated alignment errors that occurred during orientation recovery. Therefore, the exclusion of the errors that lead to the b-factor envelope does not affect the logarithm trend, as shown in Figure 3a. The numerical interpolation during structure reconstruction could not be avoided, and this might be the source of the logarithm trend in the absence of other sources. Another possibility is that the logarithm trend is due to the way information embedded in the single particle imaging dataset, where the low-resolution information is overly redundant, while the high-resolution information barely reaches the signal-to-noise threshold.

In the 2D cases, *n* measurements of the same image will boost the signal-to-noise ratio *n* times if the noise types and levels are the same for all measurements (Penczek 2010). In the 3D model reconstruction from 2D projection images, we observed the same relationship (see Figure 3). To simplify the GN noise model by combining the number of projections and the signal-noise-ratio, a new parameter, the 'effective number of projections', $N_e$, was defined as the product of the projection number and the average SNR of the individual projection image (see Equation 4). The noise term is absorbed into this new parameter $N_e$, and equation (2) becomes:

$$N_e(k) = A_{GN} k e^{Bk^2}, N_e = \frac{N}{\sigma^2} \tag{4}$$

This relation is verified with the simulation results summarized in Figure 3. In Figure 3a, the SNR values were treated as a separate parameter, independently from N (the number of measurements) and the two sets of data with different SNR values were fitted to two equations (the two black curves). In Figure 3b, the data points were merged using the effective number of projections ($N_e$). Subsequently, a single equation is adequate to describe the relationship between the resolution and the number of 'effective' projections. Using the nonlinear curve fitting algorithm, the values of coefficient $A_{GN}$ and B are 230.4 and 129.6(Å$^2$), respectively, for the GroEL simulation data.

### 3.2. Thermal fluctuation effect

Biological macromolecules exist in a thermal environment, and the structure fluctuates around the native states. In many cases, due to the functionality, molecules exist in several meta-stable conformations (Dashti et al. 2014; Chen et al. 2016; Dashti et al. 2017). Using the normal perturbation mode approach, we attempted to simulate this effect and study its influence on the achievable molecular structure resolutions at various conformation heterogeneities. The TF model described by equation (3) captures the relationship between the resolution and the number of projections in the presence of thermal fluctuations (i.e., structure heterogeneity), similar to the temperature factor in X-ray crystallography (Trueblood. et al. 1996). Compared to the Gaussian noise model, the TF model predicts a different behavior in the resolution changes. The resolution gets worse faster for molecules with larger thermal fluctuations.

In reality, both the experimental noise and molecular thermal fluctuation have impacts on the CryoEM experimental data. Therefore, it is necessary to develop a model that combines the GN and TF models. Intuitively, the following formulation is devised by treating the Gaussian noise and thermal fluctuation as independent factors that affect the structure resolutions:

$$N_e(k) = A_{TF} k e^{(Bk^2 + C\langle RMSD^2\rangle k^2)}, N_e = \frac{N}{\sigma^2} \tag{5}$$

Note the effective number of particles is used here. The fitting of the nonlinear curve to the data shown in Figure 4b yielded the $A_{TF}$, B and C parameter values ($A_{TF}$=1717, B=104.3 Å$^2$,

C=1.045). We should note that these three parameters are used to explain all the data shown in Figure 4b. Without the thermal fluctuation term $C*<RMSD^2>$ in equation (5), we will need five sets of parameters to fit the data.

### 3.3. Defocus effect on resolutions

The data collection with a CryoEM instrument is carried out at various defocus levels in a range to compensate for the information lost due to fluctuations in the contrast transfer functions. High defocus data have a strong contrast from the background for particle picking, but the information at high Fourier frequency (i.e., high resolution) is weaker. On the other hand, low defocus data preserves high-resolution information with reduced contrast from the background, making them difficult to distinguish from the background. We conducted simulations to study the effects on reconstruction resolution with several datasets, each being collected at different defocus ranges. As shown in Figure 5, model reconstruction with the low defocus data can reach higher resolutions than that with high defocus data. By increasing the ratio of low defocus data, we observed a monotonous trend. For a fixed number of projections, more low defocus data resulted improved resolutions in the reconstructed model. Furthermore, for a fixed number of projection images collected at low defocus, the model resolution does not necessarily improve by including more the high defocus data. For example, using 3,000 projection images with low defocus levels as the bottom line (green curve in Figure 5), the resolution stays approximately 3.5 Å, regardless of the number of high defocused measurements. Similarly, using 4,000 (cyan) and 10,000 (brown) projections at low defocus levels, the resolution did not improve by including images simulated with high defocus levels. Instead, the resolution got worse when including projection images at high defocus levels, especially for cases with 10,000 low defocused images. This suggests that the resolution is strongly related to the microscope defocus levels and that using low defocused data for model reconstruction will improve the final structure resolution. In practice, the signal-to-noise ratio is lower for low defocused data, making particle picking and orientation recovery more difficult. Therefore, a certain portion of high defocused data must be included during orientation recovery and model reconstruction. This suggests that high defocused data mainly contribute to initial model construction, and low defocused data contain major information to get higher resolutions.

### 3.4. Statistics from EMDB database

To compare our theory and simulation results with actual data and cross-validate the conclusions, we conducted a systematic survey on the structures determined using CryoEM single particle imaging technology, which are deposited in the EMDB database. It is worthwhile to note that only the subset of structures that met the criteria described in the Methods section was used for the statistics. The resolution distribution nicely resembles the relationship described using the noise model and the thermal fluctuation model. Using these parameters ($A_{GN}$, $A_{TF}$, B, C in equations 3 & 4), we can draw theoretical guidelines to estimate the required number of projections and obtain the desired resolutions; see Figure 6. Because the original model 1XCK weighed approximately 0.81 MDa, the molecule weight ranged from 0.5 MDa to 1 MDa compared with the analytical curves of the two models with $\langle RMSD^2 \rangle = 10$ Å$^2$ and SNR = 0.5. We observed that the resolution hardly broke the barrier of approximately 3 Å, although the number of projections covers a broad range from 20 k to 200 k. This is likely due to the thermal fluctuations or the coexistence of multiple conformations of the molecule.

### 4. Discussion and Conclusion

We studied four determinant factors that affect the reconstructed model resolutions in Cryo-EM single particle imaging technology. Based on theoretical frameworks, we systematically investigated the achievable structure resolutions and factors using numerical simulation methods, including the number of measured projections, signal-to-noise ratio, microscope defocus levels, and the heterogeneous conformations of the molecules. Two models were proposed to describe the relationship between these key factors and the model resolutions. The Gaussian Noise model is essentially the same as the formula proposed by Rosenthal and Henderson, and we found that the noise term could be combined with the number of projections by defining an 'effective number of projections'. In the thermal fluctuation model, the intrinsic dynamic characters of the sample are considered, and the final resolution is affected by the fluctuation levels. We also found that the final resolution depends on the information embedded in the data with low defocus and that adding more high defocus data does not necessarily improve the model resolution. The noise and thermal fluctuation models can be used to provide guidelines to estimate the required number of projections to reach the desired resolutions, which was validated using the statistics from the structures that were experimentally resolved using CryoEM.

Simulation studies were carried out in a tightly controlled manner so that the influence of one factor can be decoupled from that of other factors. Uniform orientation sampling is an ideal situation because orientation bias often exists in experimental datasets. In extreme cases, the missing cone problem can result in strong artifacts in the reconstructed models. Therefore, the FSC criteria measured the model consistency and not necessarily the correctness. Because of the scope of this study, we used the uniform distribution of orientations and the back-projection method to ensure that the model reconstructions were carried out properly. These operations are useful to secure the validity of the FSC criteria in the resolution cutoff. Nonetheless, the correctness of the model should be checked using complementary methods, such as visual inspection of the density maps, local resolution estimation, or validation using biochemistry assays.

In this work, the noises were simulated from a Gaussian distribution to study their influence on the model resolutions. This largely simplifies the noise sources, where the major sources include background scattering from vitreous ice, sample drifting during measurement, misalignment in the orientations and errors introduced in the orientation discretization. Some of these errors could be mimicked in the simulation framework, such as using an envelope function with b-factors. This is beyond our focus, although it may be a subject for future study.

Despite the simplicity of the formulation, the proposed models can be useful in structure determination with CryoEM single particle imaging methods. One application is to design a data collection strategy to reach the desired resolutions. SNR can be estimated based on sample screening data, and then equations (2) and (4) can be used to estimate the number of required measurements. Although the current theoretical formulation is based on uniform distributed orientation, and the free parameters may need to be refined for other molecules (because the parameters obtained in this work are for GroEL molecules), the results presented in this paper can provide hints about the obtainable structure resolution for a given number of measured projections. The data survey on the experimental structures provides evidence that the Gaussian noise and thermal fluctuation models are applicable for estimating the resolution limits in molecular structure determination.

The thermal fluctuation model can be used to assess the structure heterogeneity by reconstructing structures with subsets of data under the homogenous conformation assumption. With the obtained structure, the same simulation study can be carried out to generate a series of curves that are associated with different structural heterogeneity (see Figure 4). Then, the resolution of the experimentally determined structure as a function of the projection number can be compared with the curves to infer the level of structure heterogeneity. If the $<RMSD^2>$ values are large, then multiple structures should be reconstructed using the 3D classification approach or a similar method(Dashti et al. 2014; Frank and Ourmazd 2016; Hosseinizadeh et al. 2017).

In summary, the resolution limiting factors in the CryoEM single particle imaging method were investigated under a theoretical framework with a numerical simulation approach. The results suggest that the resolution of the reconstructed structure strongly depends on the number of measurements, image quality, molecular flexibility, and the microscope defocus levels. The results can provide guidance to design appropriate experimental strategies in data collection in general and model reconstruction in the case of molecules with heterogeneous conformations.

**Acknowledgements**

We thank Y.D. Mao, J.Y. Wu, Y.N. Zhu for helpful discussions. The computation was performed in part using the high-performance computational platform at the Peking-Tsinghua Center for Life Science at Peking University, Beijing, China.

**Funding sources**

This work was supported by NSFC (11774011, 11434001, U1530401, U1430237).

**Figure captions**

**Figure 1. The two models used for CryoEM single particle imaging data simulations.** (a) Gaussian Noise model: noises were added to the simulation data to study the influence of noise on model resolution. (b) Thermal fluctuation model: a structure ensemble was first generated to mimic the structure heterogeneity. Each single particle projection was simulated based on a randomly picked structure from the ensemble. Three representative models for GroEL are shown in (b): the original structure is in gray; the yellow and blue structures were generated using normal mode perturbations.

**Figure 2. The workflow for single particle imaging data stimulation.** (a) The protocol for the projection data simulation with defocused lenses and Gaussian noises from a single 3D structure. (b) The procedure for structure ensemble generation to model the heterogeneity of biomolecular conformations.

**Figure 3. Relation between model resolution and data quantity/quality. (a)** The model resolution depends on the number of measurements. Two sets of data are plotted with signal to noise ratios (SNR) of 0.2 and 0.4. The lines were obtained by fitting the data points using the Gaussian Noise model. R-square values were 0.9946 and 0.9865, respectively. **(b)** The noise term was combined with the number of projections. The two sets of data in Figure 3**a** were fitted with a single set of parameters by defining the effective number of projections. R-square of the line is 0.9782.

**Figure 4. The influence of thermal fluctuation to the structure resolution. (a)** The relationship between the resolution and number of projections at various thermal fluctuation levels. The thermal fluctuation is quantified using the average square root-mean-square-derivation $\langle RMSD^2 \rangle$ of the structure ensemble. All the projections were simulated with Gaussian noise with an SNR of 1.00. The R-square values of the fittings to the GN model are approximately 0.96. **(b)** The noise term is combined with the number of projections. The five datasets in Figure 4 were fit using a single set of parameters with an R-square of 0.9822.

**Figure 5. The influence of defocus levels on model reconstruction resolutions.** The two solid lines indicate the upper (lower) limits of model resolution using the data collected with two defocus ranges. Higher resolutions can be reached using data with a low defocus. The dashed lines show the cases with a fixed number of projections with low defocus. The green line represents the results with 3000 projections at low defocuses (1.0 μm – 1.5 μm) and various numbers of projections at high defocuses (1.5 μm – 2.0 μm). The cyan line shows the same study with 4000 low defocused projections, and the brown line represents the results with 10,000 low defocused images.

**Figure 6. The distribution of model resolutions as a function of particle numbers for experimentally determined structures.** The data points are colored based on the molecular weights of the sample molecules. The theoretical lines indicate the resolution limits for molecules with molecular weights between 0.5 and 1 MDa. The curves were drawn with SNR ≈ 0.5 for GN model, and using $<RMSD^2> = 10$ Å$^2$ for the TF model using the free parameters fit from the simulation results.

**Tables**

| Group ID | $<RMSD>$ *(Å) | $<RMSD^2>$** (Å$^2$) |
|---|---|---|
| 1 | 0.79 | 0.90 |
| 2 | 1.24 | 1.65 |
| 3 | 1.62 | 3.16 |
| 4 | 2.08 | 5.86 |
| 5 | 2.52 | 10.78 |
| 6 | 3.02 | 17.35 |
| 7 | 3.54 | 23.03 |
| 8 | 4.02 | 27.88 |
| 9 | 4.53 | 32.51 |
| 10 | 4.99 | 35.80 |

Table 1. The characteristics of the structure ensembles after incorporating the thermal fluctuations. In each group, there were 100 structures selected using the protocol described in the main text.

*<RMSD>: average RMSD with respect to the original model.

**<RMSD$^2$>: The mean square of the pairwise RMSD within the ensemble.

| Models<br>Factors | Gaussian noise model | Thermal fluctuation model | De-focus effect |
|---|---|---|---|
| Conformation Heterogeneity | No | Yes (control) | No |
| CTF effect | Yes | Yes | Yes (control) |
| Noise level | Yes (control) | Yes (control) | Yes |

Table 2. The resolution determinant factors explored in different models.

**Figure 1.**

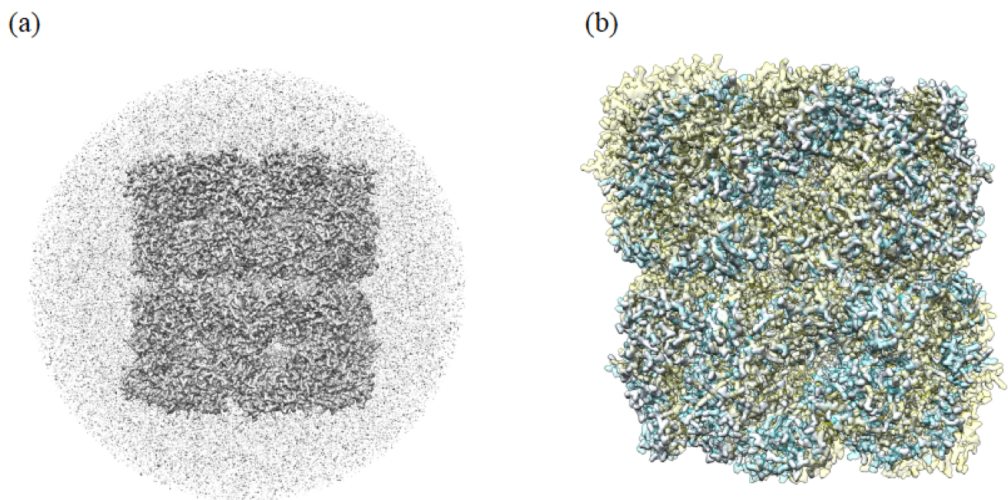

**Figure 2.**

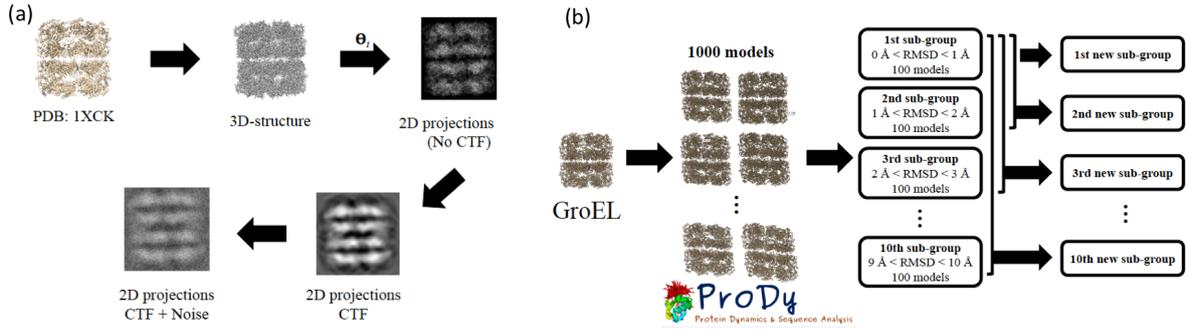

**Figure 3.**

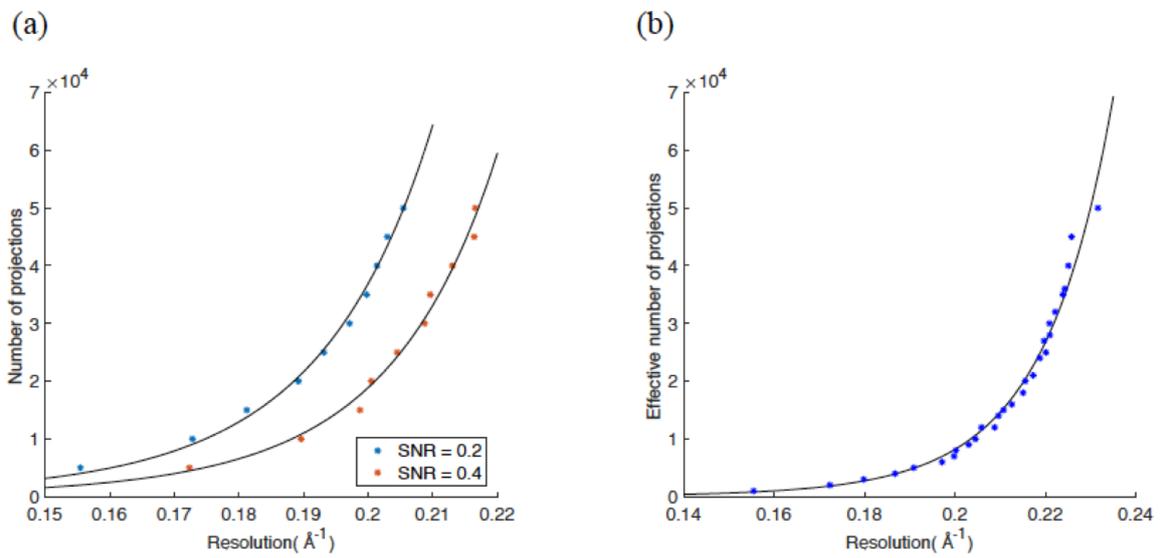

**Figure 4.**

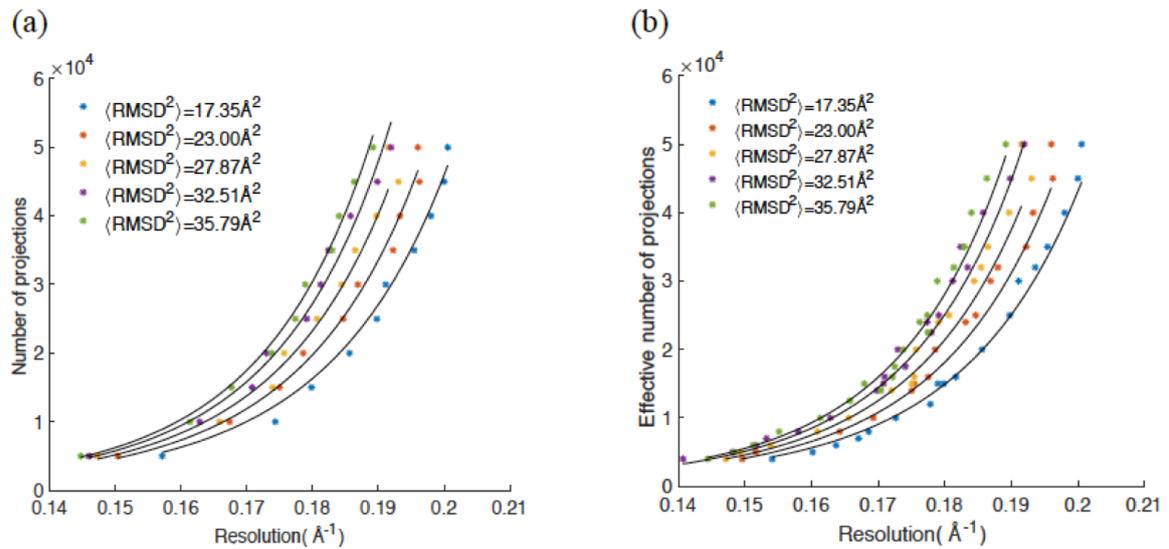

**Figure 5.**

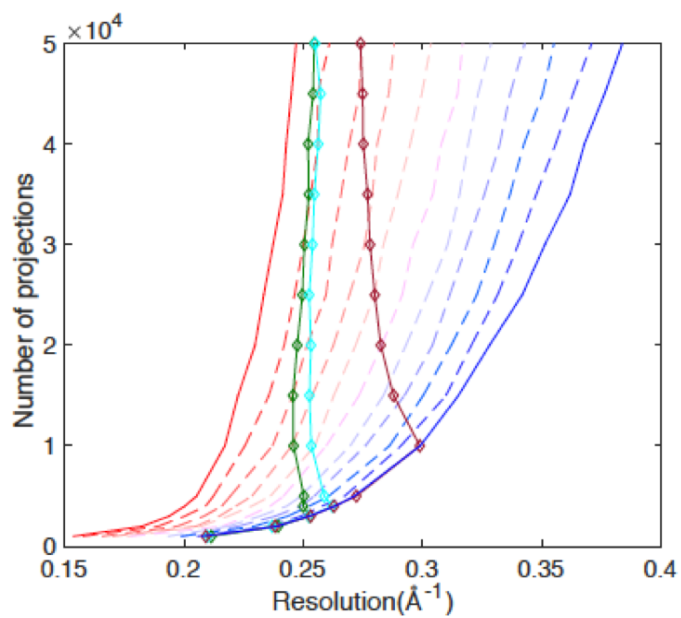

**Figure 6.**

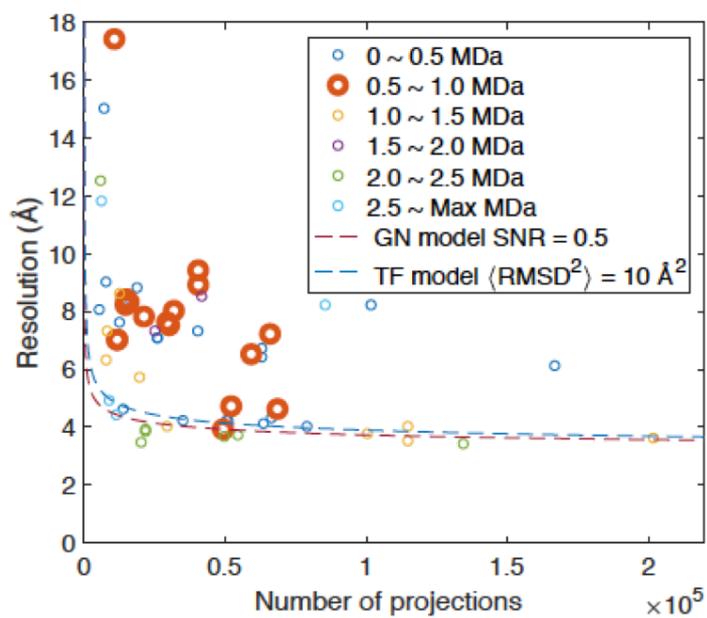